\documentclass{PoS}
\usepackage{epsf}
\usepackage[dvips]{epsfig}

\PoS{PoS(BDMH2004)101}

\title{How star clusters could 
survive low star formation efficiencies}

\ShortTitle{Star Clusters surviving low SFE}

\author{\speaker{Michael Fellhauer}\\
        Sternwarte, Univerity of Bonn, Germany\\
        E-mail: \email{mike@astro.uni-bonn.de}}

\author{Pavel Kroupa\\
        Sternwarte, University of Bonn, Germany\\
        E-mail: \email{pavel@astro.uni-bonn.de}}

\abstract{
  After the stars of a new, embedded star cluster have formed they
  blow the remaining gas out of the cluster.  Especially winds of high
  mass stars and definitely the on-set of the first super novae can
  remove the residual gas from a cluster.  This leads to a very violent
  mass-loss and leaves the cluster out of virial equilibrium.
  Standard models predict that the star formation efficiency (SFE) has
  to be about $33$~per cent for sudden (within one crossing-time of the
  cluster) gas expulsion to retain some of the stars in a bound
  cluster.  If the efficiency is lower the stars of the cluster
  disperse completely.  

  Recent observations reveal that in strong star bursts star clusters
  do not form in isolation but in complexes containing dozens and up
  to several hundred star clusters (super-clusters).  By carrying out
  numerical experiments we demonstrate that in these environments (i.e.\
  the deeper potential of the star cluster complex and the merging
  process of the star clusters within these super-clusters) 
  the SFEs could be as low as $20$~per cent, leaving a gravitationally
  bound stellar population.  We demonstrate that the merging of the
  first clusters happens faster than the dissolution time therefore
  enabling more stars to stay bound within the merger object.

  Such an object resembles the outer Milky
  Way globular clusters and the faint fuzzy star clusters recently
  discovered in NGC~1023. 
}

\FullConference{Baryons in Dark Matter Halos\\
		 5-9 October 2004\\
		 Novigrad, Croatia}

\begin{document}

\section{Results}

\begin{figure}[h!]
  \centering
  \epsfxsize=6.5cm
  \epsfysize=6.5cm
  \epsffile{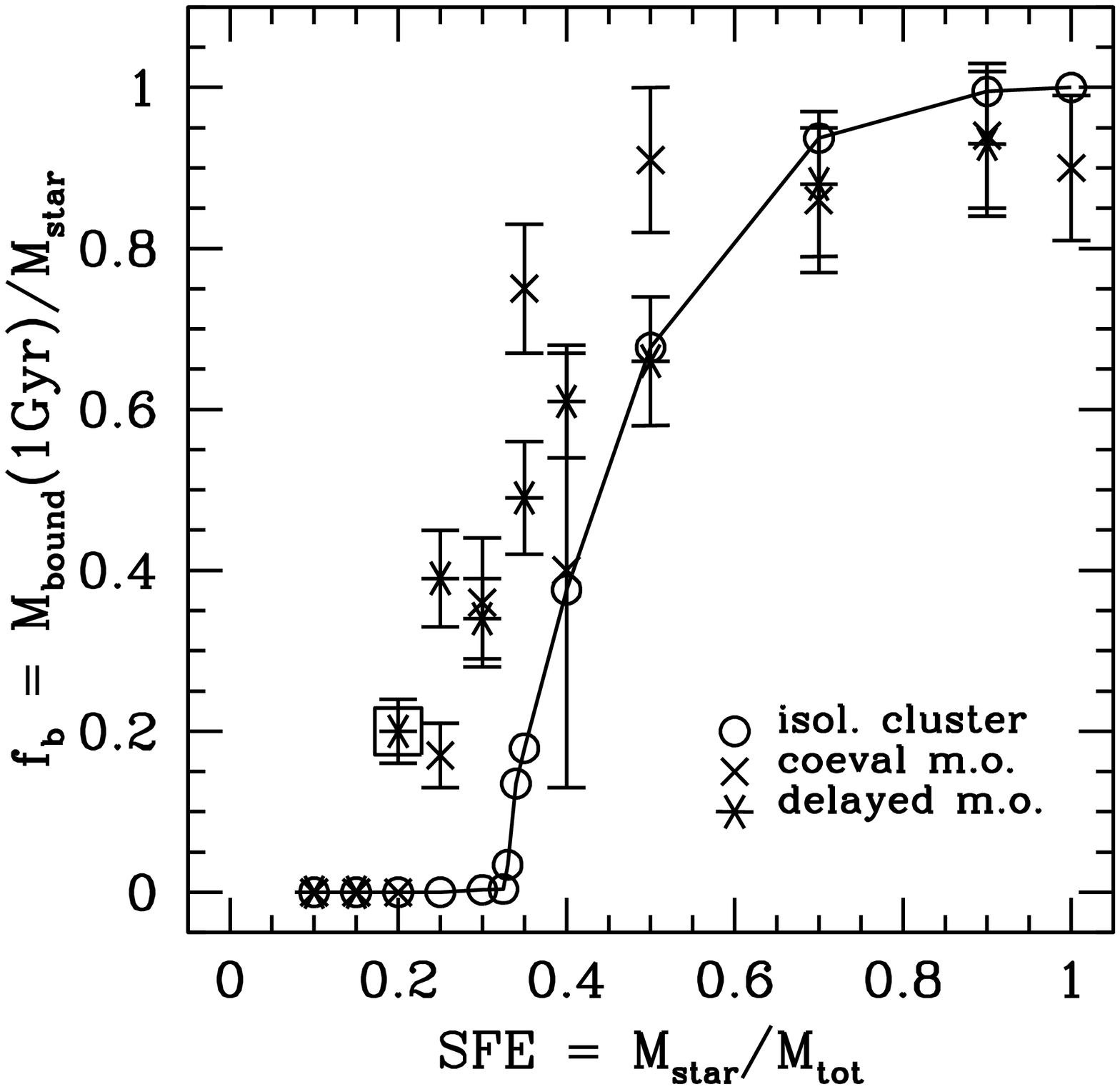}
  \epsfxsize=6.5cm
  \epsfysize=6.5cm
  \epsffile{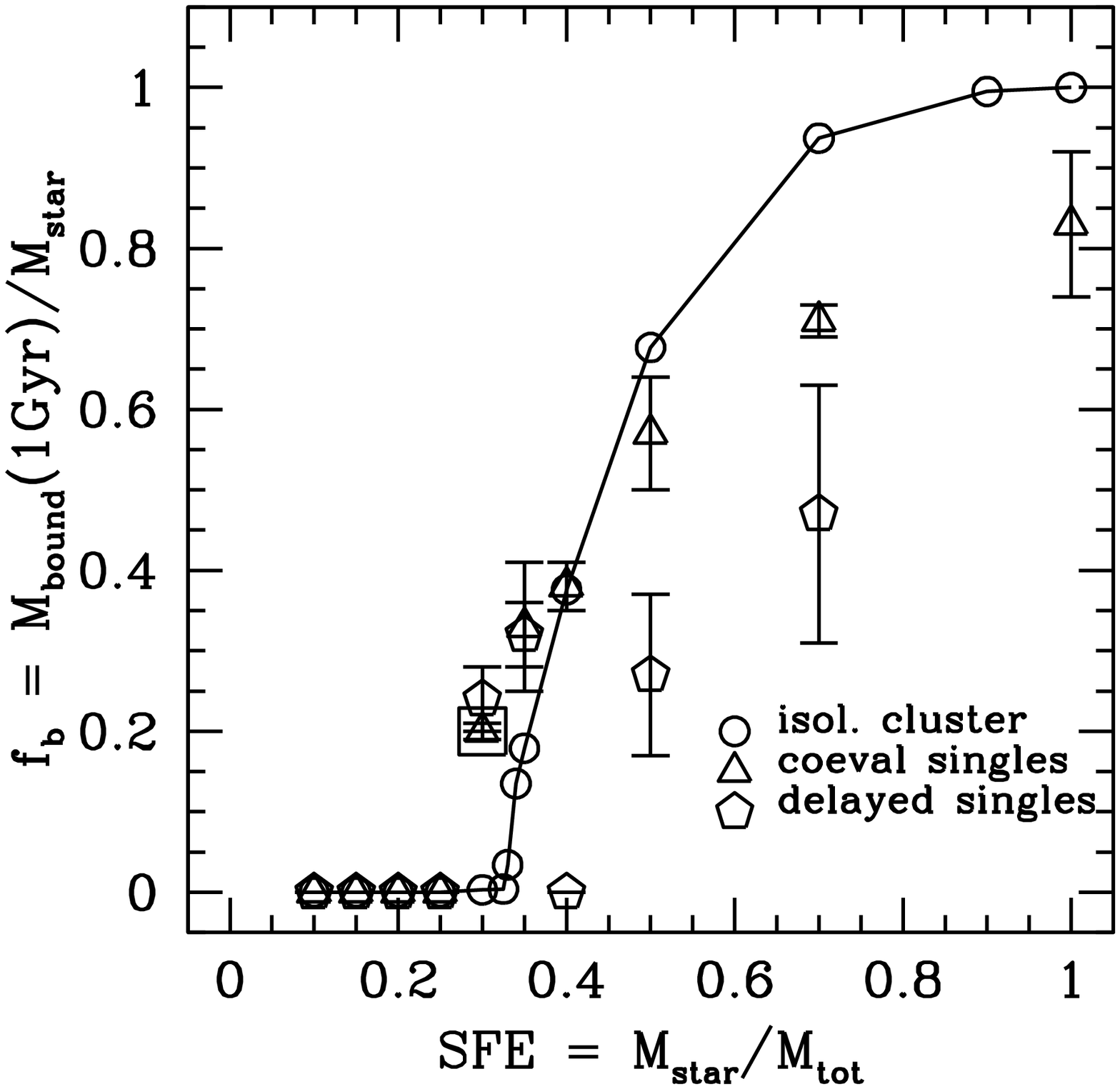}
  \caption{
    The fraction $f_{b}$ of bound particles (stars) after $1$~Gyr of
    evolution is plotted versus the the star formation efficiency
    (SFE), i.e.\ the fraction of mass in the embedded cluster which is
    transformed into stars.  The remaining gas is blown out within a
    cluster crossing time.  
    Large open circles show the results of the isolated clusters.
    These results agree with previous results in the literature.  
    The star cluster complex contains $20$ star clusters, each
    modelled as a Plummer sphere with $100000$~particles, and is
    modelled as a Plummer distribution, with the star clusters as
    'particles'.  This Plummer distribution is given a Plummer radius
    of $20$~pc, a cut-off radius of $100$~pc and a crossing time of
    $5.9$~Myr.  This is a very dense configuration which will lead to
    a fast merging of the star clusters into one massive merger
    object, and resembles some of the cluster complexes observed in
    the Antennae galaxies. 
    The mass-loss due to gas-expulsion is modelled by all particles
    loosing a fraction of their mass linearly over a crossing-time of
    the single star cluster.  We consider two mass-loss models.  In
    the coeval model every star cluster starts immediately and at the
    same time to loose mass and in the delayed model the star clusters
    start to loose their mass randomly during the first crossing time
    of the super-cluster.  
    The star cluster complex orbits circularly at a distance of
    $10$~kpc around an analytical galactic potential with a flat
    rotation curve of $220$~kms$^{-1}$.  
    Crosses are the merger objects of the simulations with coeval
    gas-expulsion, six-pointed stars denote the merger objects in the
    simulations with randomly delayed gas expulsion (left panel).
    Small open triangles and small open pentagons are surviving and
    escaped star clusters in the coeval and the delayed case
    respectively (right panel).  
    Almost all clusters which do not end up inside the merger object
    have a smaller bound mass than in the isolated case.  There are
    even star clusters dissolving completely when the SFE is $70$~per
    cent.  On the other hand there are rare cases where single
    clusters escape and survive even at low star formation
    efficiencies.  Two single clusters escape and survive the coeval
    simulation with a SFE of $30$~per cent.  
    The building up of a merger object with its deeper potential well
    favours the survival of a bound object that retains more of its
    stars than an isolated single cluster would, as long as the SFE is
    below $60$~per cent.  If the SFE is higher destructive processes
    during the merging process lead to a mass-loss, i.e.\ stars that
    are expelled as a result of the kinetic energy surplus produced
    during the merging of the clusters (the stars are then found in
    the tidal tails), leaving the remaining object with a smaller
    bound mass fraction than the star clusters would have had if they
    would have formed in isolation.
    However, as a major result we find that cluster formation in
    complexes allows star clusters to survive even if the SFE is as
    low as $20$~per cent.  In a dense star cluster complex the
    crossing time of the star clusters through the super-cluster, and
    therefore the merging time-scale, is short enough that some star
    clusters have already merged before they expel their gas.  The
    much deeper potential wells of these merger objects are able to
    retain the stars more effectively than isolated clusters would. 
  }
\end{figure}


\end{document}